\documentstyle[floats,multicol,aps,epsf,prb]{revtex}

\begin{document}
\draft
\twocolumn[\hsize\textwidth\columnwidth\hsize\csname @twocolumnfalse\endcsname
\title{
Simple description of the anisotropic two-channel Kondo problem
}
\author{P. Coleman and A. J. Schofield}
\address{Department of Physics, Rutgers University,
Piscataway, P.O. Box 849, NJ--08855--0849}
\date{\today}
\maketitle
\begin{abstract}
We adapt strong-coupling methods first used in the 
one-channel Kondo model to develop a simple description of the 
spin-$1\over 2$ two-channel Kondo model with
channel anisotropy. Our method exploits spin-charge decoupling to
develop a compactified Hamiltonian that describes
the spin excitations.  The structure of the
fixed-point Hamiltonian and quasiparticle impurity S-matrix
are incompatible with a Fermi liquid description.
\end{abstract}
\vskip 0.2 truein
\vskip2pc]

An important question in the current debate on non-Fermi liquids
concerns their stability in a real-world environment.  Though it is
possible to construct models with non-Fermi liquid ground-states,
conventional wisdom holds that real-world perturbations absent from
the model will drive a non-Fermi liquid back to a Fermi liquid.  This
has led to controversy in connection with two non-Fermi liquid models:
the one-dimensional Luttinger liquid~\cite{haldane81,pwa90}, and the
single impurity two-channel Kondo model, where the contentious
perturbations are dimensionality~\cite{finkelstein}, and channel
asymmetry respectively.

In this Letter we confront this issue in the context of the
asymmetric two-channel Kondo model.  This model was first
introduced by Nozi\` eres and Blandin~\cite{nozieres80}
\begin{eqnarray}
{\cal H}'  =&\displaystyle
it\sum_{n \lambda \sigma}[c^\dagger_{\lambda\sigma}(n+1)
c^{}_{\lambda\sigma}(n) - {\rm H.c}.]  \nonumber \\ 
&\displaystyle + 
[J_1\vec \sigma_1(0) + J_2\vec \sigma_2(0)] \cdot \vec S_d,
\label{H'} 
\end{eqnarray}
where $\lambda=1,2$ labels two independent one-dimensional conducting
chains, $\vec\sigma_{\lambda}(0) = c^{\dagger}_{\lambda\alpha}(0)
\vec\sigma^{}_{\alpha\beta} c^{}_{\lambda\beta}(0)$
is their spin density at the origin and $\vec S_d$ is a localized spin
$1/2$ operator.  By tuning the anisotropy $J_2/J_1$, this model
interpolates continuously between a non-Fermi liquid state at the
channel-isotropic point~\cite{andrei84,tsvelik84} ($J_2/J_1 =1$) and a
Fermi liquid in the single channel limit~\cite{nozieres74}
($J_2/J_1=0,\infty$).  Fabrizio, Gogolin and
Nozi\`eres~\cite{gogolin95} have recently argued that the relevant
perturbation of channel anisotropy immediately restores Fermi liquid
behavior.  However, when we contrast this model with the corresponding
spin-$1$ model, itself a well-established Fermi
liquid~\cite{nozieres80,hewson93}, we are faced with a series of
puzzling differences.  A new Bethe Ansatz solution\cite{andrei95}
shows clear qualitative differences between the excitation spectra of
these two models. In particular a new type of singlet excitation
present in the spin 1/2 model is absent in its spin 1 counterpart.  In
addition, ertain features of a two-channel Fermi liquid close to
channel isotropy are expected to be universal~\cite{nozieres80}. Most
notably, channel symmetry is expected to constrain the ratio of inter-
and intra-channel interactions, leading to a Wilson ratio close to
$8/3$.  By contrast, the $S=1/2$ model exhibits vanishingly small
Wilson ratios in this region,\cite{gogolin95,andrei95} which require
the introduction of an ad-hoc inter-chain Fermi liquid interaction of
ever increasing size\cite{gogolin95}.

\begin{figure}[btp]
\epsfxsize=3.32in \epsfbox{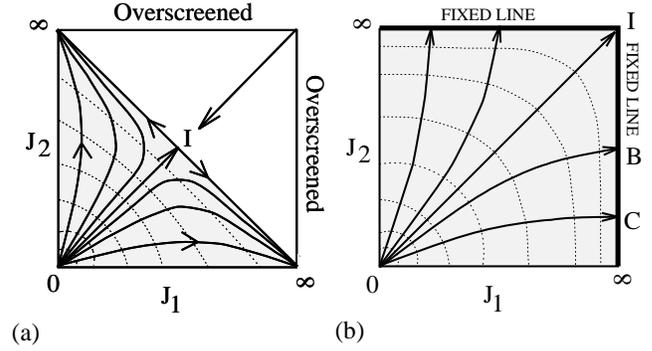}
\protect\caption{(a) Scaling trajectories for
the two-channel Kondo model showing intermediate
coupling fixed point and separatrix; (b) scaling trajectories
of the compactified two-channel Kondo model. The beta
function for the two procedures coincides at weak coupling,
but in (b) the elimination of the  over-screened fixed point
leads to a redefinition of the coupling constants at strong-coupling
that shifts the separatrix linking to infinite coupling.
}
\label{fig1}
\end{figure}
We clarify these points here by using a strong-coupling expansion.
A novel approach is required, because the scaling  physics of the
two-channel $S=1/2$ Kondo model is controlled by a separatrix at
{\em intermediate} rather than infinite
coupling (see Fig. 1).  To overcome this
difficulty, we exploit spin-charge decoupling
to remove the 
charge degrees of freedom~\cite{coleman95}.  
This procedure eliminates the 
over-screened fixed points, (Fig.~\ref{fig1}) moving the attractive
fixed points to infinity where a strong-coupling expansion
can be made.  In the fixed point Hamiltonian we 
derive, we 
find that  the spin-quasiparticles occur in a  $S=0$
or a triplet $S=1$ branch. Unlike the Fermi liquid description,
these quasiparticles
are intrinsically channel
off-diagonal and  their interactions are
completely constrained in terms of the impurity phase shifts in such
a way that the Wilson ratio drops to 
zero in the vicinity of channel isotropy. These results 
substantiate a conclusion that channel
asymmetry does not restore a Fermi liquid.

The method used exploits two observations: (i) that the charge and pair
degrees of freedom can be combined into a single $SU(2)$ spin operator,
commonly called  ``isospin'',\cite{pwa58,nambu60} and (ii) that 
the spin and isospin variables
of a single linearized conduction chain form two independent spin degrees
of freedom. For a linearized one-dimensional band, the mapping
\begin{equation}
\begin{array}{rl}
\vec \sigma_1 (x) &\displaystyle\longleftrightarrow \vec \sigma(x),
\nonumber \\
\vec \sigma_2 (x) &\displaystyle\longleftrightarrow \vec \tau  (x), 
\label{map} \\
\end{array}
\end{equation}
where
\begin{eqnarray}
\vec{\sigma_1}&=&
(c^{\dagger}_{\uparrow},\ c^{\dagger}_{\downarrow}) \cdot
\vec{\sigma} \cdot \left( \begin{array}{c} c^{}_\uparrow \\
c^{}_{\downarrow} \end{array} \right) \; ,\\
\vec\tau&=&
(c^{\dagger}_{\uparrow}, \; c^{}_{\downarrow}) \cdot
\vec{\sigma} \cdot \left( \begin{array}{c} c^{}_\uparrow \\
c^{\dagger}_{\downarrow} \end{array} \right) \; 
\end{eqnarray}
are the spin and isospin density respectively,
preserves the spin operator algebra. Under this mapping we may remove
the (inert) charge degrees of freedom of the original two-channel
Kondo model and compactify the spin-physics of this model into the
following Hamiltonian
\begin{equation}
{\cal H} = {\cal H}_{\rm band}+{\cal H}_{\rm int} \; .
\label{H}
\end{equation}
Here
\begin{eqnarray}
{\cal H}_{\rm band}&=&it\sum_{n}
[c^\dagger_{\sigma}(n+1)c_{\sigma}(n) - {\rm H.c.}] \; , \\
{\cal H}_{\rm int}&=&  [J_1\vec \sigma(0) 
+J_2\vec \tau(0)] \cdot \vec S_d
\end{eqnarray}
and the spin and isospin of the single band 
represent respectively the spin-density of channel one
and two in Eq.~(\ref{H'}).  
The advantage of this procedure derives from
the inability of spin and isospin to co-exist at a single site on a
lattice, preventing over-screening.  The unstable over-screened fixed
point is thereby removed from the scaling trajectories and the scaling
trajectories are deformed at strong coupling so that the
(non-universal) location of the non-Fermi liquid fixed point is shifted to infinite
coupling (Fig.~\ref{fig1}).

For the conventional Kondo model, our ability to identify the
ground-state as a Fermi-liquid is due to the special duality
between the Anderson and Kondo models.  Under this duality, the strong
coupling limit of each model may be canonically transformed into the
weak-coupling limit of its dual counterpart. These dual models
represent two extreme limits of a {\em single} scaling trajectory.
Weak coupling in the Kondo model controls the high temperature
local-moment physics~\cite{schrieffer66}; strong-coupling in the Kondo
model controls the low-temperature physics. Duality permits us to map
the strong coupling physics of the Kondo model onto an Anderson model
at weak coupling with renormalized parameters~\cite{hewson93}. This
provides the basis of Nozi\`eres' phase shift Fermi liquid description
of the Kondo model~\cite{nozieres74}.

We now construct the dual to the compactified 
two-channel Kondo model of Eq.~(\ref{H}).  The natural
language here is that of Majorana fermions, whereby the two Fermi fields
are rewritten in terms of four real (Majorana) components $\Psi_0$
and $\vec{\Psi}$ as follows
\begin{equation}
\left(\begin{array}{c} c_{\uparrow}(n)\\c_{\downarrow}(n)
      \end{array} \right)
= {1 \over \sqrt{2}}\left( \Psi_0(n) + i \vec\Psi(n)\cdot  \vec\sigma
        \right) \left( \begin{array}{l} 0 \\ -i \end{array} \right).
\label{c}
\end{equation}
In this representation we may write
\begin{eqnarray}
{\cal H}_{\rm band}
&=&it \sum_{n, \alpha =0}^3 \Psi_{\alpha}(n+1)\Psi_{\alpha}(n) \; ,
\\
{\cal H}_{\rm int}&=& i\left[ J \vec \Psi(0) \times
\vec \Psi(0)
+ 2 A \Psi_0(0) \vec \Psi(0)\right]\cdot \vec S_d \; ,
\label{facts}
\end{eqnarray}
where $ J=\frac{1}{2}(J_1+J_2)$ and $A=\frac{1}{2}(J_1-J_2)$.  Thus
scalar and vector components of the conduction sea are completely
decoupled in the channel symmetric model ($A=0$).  By contrast, in the
one-channel limit, (${A \over J} = \pm 1$) all four components of the
conduction Majoranas are symmetrically coupled to the local moment,
forming a model with an $O(4)\sim SU(2)\times SU(2) $ symmetry derived
from spin conservation on each separate channel.  The symmetric
Anderson model that is dual to this limit
\begin{equation}
{\cal H}_A = {\cal H}_{\rm band}
+  iV[c^\dagger_{\sigma}(0)d_{\sigma} - {\rm H.c.} ] + 
\frac{U}{2} (\hat n_{d}-1)^2 \; ,
\label{Hand}
\end{equation}
can be rewritten in the Majorana representation to display this $O(4)$
symmetry
\begin{eqnarray}
{\cal H}_A = {\cal H}_{\rm band}
+ i V \sum_{\alpha=0}^{3} \Psi_{\alpha}(0)d_{\alpha} 
- U d_{0}d_1d_2d_3 \; .
\label{Handx}
\end{eqnarray}
To develop the dual to the two-channel model
we break the $O(4)$ symmetry in the hybridization down to 
an $O(3)$ symmetry, and write the following model
\begin{equation}
{\cal H}_M = {\cal H}_{\rm band} 
+ i[ V_s \Psi_{0}(0)d_{0}+  V_v \vec \Psi(0)\cdot \vec
d]- U d_{0}d_1d_2d_3 \; .
\label{Handy}
\end{equation}
When we carry out a Schrieffer-Wolff canonical transformation that
eliminates the hybridization terms, we find that the compactified
two-channel model is recovered with
\begin{equation}
J = {2 (V_v)^2  \over U} \; , \quad \qquad
A = {2 V_s V_v \over U}\; . \label{swolff}
\end{equation}
Once we can confirm that the compactified two-channel Kondo model
scales to strong coupling, we may immediately use the Majorana
resonant level model to describe the low-temperature physics.

Next, consider 
the stability of the large $J$ limit.
For
$J \gg t$, ${\cal H}_{\rm band}$ is a perturbation and 
the structure of the eigenstates is dominated by  ${\cal H}_{\rm int}$.
The eigenstates of ${\cal H}_{\rm int}$ involve 
two  singlets  and two triplets
separated by an energy of order $J$.  To examine the
stability of this limit, we systematically
develop a $t/J$ expansion. This is done by using
a canonical transformation which removes, in powers
of $t/J$, the mixing between the singlet and triplet
subspaces.
Working to order $1/J^2$ it is sufficient 
to consider only the hopping between
site $0$ and site $1$ to obtain the  strong coupling
Hamiltonian
\begin{eqnarray}
\tilde{\cal H}^*&=& \! \!
it\sum_{n>0} \vec \Psi_{\alpha}(n+1)\cdot \vec \Psi(n)
+ it\sum_{n\ge 0} \Psi_{0}(n+1) \Psi_{0}(n)\nonumber\\
&+& 
iV_s^*\Psi_0(0)\Phi- U^*
\Phi
\Psi_1(1) \Psi_2(1) \Psi_3(1) \; ,
\label{eff} 
\end{eqnarray}
where $U^*= {3 t^3 / 4J^2}$, $V_s^*=A(1-t^2/2J)$ and $\Phi=-2i
\Psi_1(0) \Psi_2(0) \Psi_3(0)$ is a single Majorana field that is
hybridized with the scalar fermions.  By hybridizing with a
zero-energy fermion, the scalar fermions experience unitary
scattering, with phase-shift $\delta_o= \pi/2$.  Since the vector
fermions are excluded from the origin, they also experience unitary
scattering, but with a resonance width of order $t$.  The form of this
Hamiltonian is thus isomorphic with ${\cal H}_M$ given above and since
$U^*\ll t$, the model is weak-coupling.

The isotropic case $A=0$ has been studied in detail in previous
papers~\cite{emery92,sengupta94,coleman95}.  At this point the scalar
fermions decouple to form a free band of unscattered singlet
excitations.  The coupling $U^*$ between the localized Majorana $\Phi$
is a `dangerous irrelevant variable'. Though its vertex corrections
can be neglected, it gives rise to logarithmically singular
contributions to the electron self-energy and thermodynamics. At
$T=0$, $\Phi$ is asymptotically decoupled from the Fermi sea, forming
a fermionic zero mode. This is the feature that is
responsible\cite{emery92,sengupta94} for the fractional zero-point
entropy ${1\over 2} \ln 2$~\cite{andrei84,tsvelik84}.  At finite
temperatures the three-body interaction with the conduction sea can be
treated perturbatively.  The `dangerous' terms in the free energy to
lead to a logarithmic temperature dependence of the spin
susceptibility $\chi_{\sigma}\sim \ln T$ and specific heat
$\gamma=\frac{{\rm C}_v}{T}\sim \ln T$ which  combine to give 
a dimensionless Wilson ratio of 8/3.

In the channel anisotropic case $A \neq 0$, the fermion $\Phi$
acquires a finite lifetime: $\tau_{\Phi}^{-1}=A^2/2t$.  Processes at
energies $\omega \gtrsim \tau_{\Phi}^{-1}$ are insensitive to the
anisotropy.  Marginal spin-physics is thus exhibited over a finite
temperature range $\tau_{\Phi}^{-1} \ll T \ll t$.  Once $T \ll
(\tau_{\Phi}^{-1})$ the correct description will then be in terms of
an effective Hamiltonian in which only the lowest lying singlet state
remains. Since there is no residual degeneracy, this fixed point is
stable.  The scattering and interactions of the vector and scalar
Majorana fermions off the singlet state will then be described by a
renormalized Majorana resonant level model, of the type given above
\begin{eqnarray}
\tilde {\cal H}_{M}&=&{\cal H}_M[V_s^*, V_v^*, U^*]
\nonumber \\
&+& igB\left( d_1 d_2 - d_0 d_3 \right)
+i\mu\left( d_1 d_2 + d_0 d_3 \right) \; ,
\end{eqnarray}
where we have explicitly included a chemical potential and magnetic
field in the $z$ direction.

We now use
renormalized perturbation theory about $U^*=0$ point~\cite{hewson93} to obtain the
low-temperature thermodynamics.  We work to leading order in $U^*$ and
consider the large bandwidth limit, with a density of states $\rho$.
The scalar and vector fermions develop resonant levels of width
$\Delta_s=\pi \rho |V^*_s|^2$ and $\Delta_v= \pi \rho |V^*_v|^2$
respectively.  The linear specific heat coefficient is then given by
$\gamma_{\rm imp}=\frac{\pi^2}{3} \rho_{\rm imp}$, where
\begin{equation}
\rho_{\rm imp}  = {1 \over 2\pi}\left({3 \over \Delta_v} + {1\over
\Delta_s} \right) \; .\label{gamma}
\end{equation}
\begin{figure}[btp]
\epsfxsize=3.32in \epsfbox{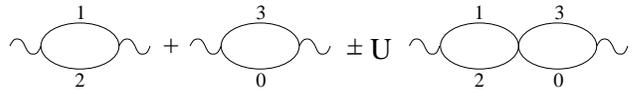}
\caption{Leading terms in the charge and spin
susceptibilities of the impurity.
Positive and negative signs are for the charge and spin susceptibility
respectively.}
\label{sus}
\end{figure}
Fig.~\ref{sus} shows the diagrams that need to be calculated to
determine the susceptibilities. These lead to
\begin{eqnarray}
\chi_c^{\rm imp}&=&{1\over \pi \Delta_v}+{\ln (\Delta_v/\Delta_s) \over \pi
(\Delta_v-\Delta_s)} \left( 1 - {2 {U^*}\over \pi \Delta_v}\right),
\label{charge}
\\
\chi_\sigma^{\rm imp}
&=&{1\over \pi \Delta_v}+{\ln (\Delta_v/\Delta_s) \over \pi
(\Delta_v-\Delta_s)} \left(1 + {2 {U^*}\over \pi \Delta_v}\right).
\label{spin}
\end{eqnarray}
Since the two-channel Kondo model corresponds to
the large $U$ limit of ${\cal H}_M$, it follows that
the renormalized calculation must lead to
a vanishing charge susceptibility $\chi_c^{\rm imp}=0$.
Using (\ref{charge}) to eliminate $U^*$, it follows that
\begin{equation}
\chi_\sigma^{\rm imp}={2 \over \pi} 
\left[ {1\over \Delta_v} + {\ln (\Delta_s
/\Delta_v) \over (\Delta_s-\Delta_v)} \right] \;. \label{spin2}
\end{equation}
From (\ref{spin2}) and (\ref{gamma}) we  obtain the Wilson ratio
\begin{equation}
R={\pi^2  \over 3 }{\chi_\sigma^{\rm imp} \over \gamma_{\rm
imp}} = 4 \left[{1 + (\ln \alpha)/(\alpha-1) \over 3 + 1/\alpha} \right] \; ,
\label{wilson}
\end{equation}
where $\alpha=\Delta_s /\Delta_v$ (see Fig.~\ref{wilsonfig}). 
This result extrapolates between
$-4\alpha\ln \alpha$ at low anisotropy and $2$ as
$\alpha \rightarrow 1$ (the single channel limit), in qualitative
agreement with the Bethe Ansatz solution.~\cite{andrei95}  At small anisotropy, the physics
is dominated by the spinless scalar fermions, and this is the origin
of the small spin-susceptibility and Wilson ratio.  In the
thermodynamic Bethe Ansatz solution, there is a two-stage quenching of
the impurity spin entropy: below $T_K$ the entropy saturates at ${1
\over 2} \ln 2$ before quenching to $0$ at a new low
energy scale. We may identify this second scale with
$\tau_{\Phi}^{-1}$.  In the isotropic limit, the Wilson ratio found by
Bethe Ansatz goes vanishes as
$R \sim -\Delta \ln \Delta$ (where $\Delta$ is the ratio
of the two energy scales) in exactly the same fashion
as the calculation presented above.
\begin{figure}[btp]
\epsfxsize=3.32in \epsfbox{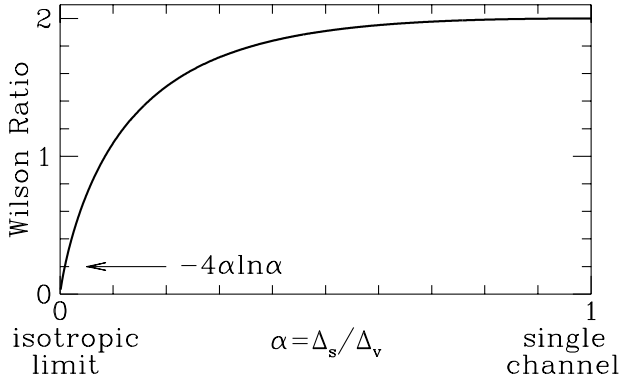}
\protect\caption{The dimensionless Wilson ratio, $\chi^{\rm
imp}_\sigma/\gamma_{\rm imp}$, as a function of the anisotropy
$\alpha=\Delta_s/\Delta_v$ [see Eq.~(\ref{wilson})].}
\label{wilsonfig}
\end{figure}

We now turn to the question of whether this state is a Fermi liquid.
Using Eq.~(\ref{map}) we identify the total spin of the excitations as 
$\vec{S}=(\vec{\sigma}+\vec{\tau})/2$.
At the fixed point, the spin-excitations in the two channels combine
into two types of spin excitation, a singlet and a triplet represented
by the scalar and vector Majorana fermions:
\begin{eqnarray}
S&=&0  \quad \quad \Psi^{\dagger}_0(k)|0\rangle \; , \\
S&=&1 \quad  \left\{ 
\begin{array}{lc}
S_z=\pm 1 & {1\over \sqrt{2}}
[\Psi^{\dagger}_1(k)\pm i\Psi^{\dagger}_2(k)]|0\rangle
\\
S_z=0 &  \Psi^{\dagger}_3(k)|0\rangle
\end{array}  
\right. \; ,
\end{eqnarray}
where $0<k<\pi$. These excitations
are formed by correlating the spins across the two channels---a 
feature reminiscent of the `spin fusion' process that 
plays a central r\^ ole in the Bethe Ansatz solution\cite{andrei84,andrei95}.
At the isotropic point, the scalar excitations decouple from the
impurity.  This is confirmed by the Bethe Ansatz solution,
which shows that the linear specific heat capacity of the conduction
sea increases by a factor of $4/3$ in going from the isotropic to the
anisotropic solutions\cite{tsvelik}. 
From the fixed point
Hamiltonian, we deduce that the one-particle S-matrix of the spin
excitations has the form
\begin{equation}
S_{\rm 2-channel}=e^{2i\delta_s(\epsilon)} {\cal P}_s +
e^{2i\delta_v(\epsilon)} {\cal P}_v \; , \label{smatrix}
\end{equation}
where
$
\delta_{s,v}(\epsilon)={\pi\over 2} + {\rm tan^{-1}}
\left(\displaystyle \epsilon
\over \Delta_{s,v} \right) \; 
$
and 
\begin{equation}
{\cal P}_{v}= \frac{1}{4}\left[3+  
\vec \sigma \cdot \vec \tau)\right] \; \; , \; \;
{\cal P}_{s}=\frac{1}{4}\left[1-  
\vec \sigma \cdot \vec \tau)\right] \; ,
\end{equation}
project the excitations into the singlet and triplet
channel~\cite{footnote}.  
This S-matrix is explicitly channel off-diagonal. 
Were the ground-state a Fermi
liquid, the one-quasiparticle S-matrix
would be channel diagonal, with spin-${1\over 2}$ quasiparticles
that are individually confined to a definite channel. 
Unlike the
Fermi liquid description, the quasiparticle interaction is completely
constrained by the phase shifts, and does not need to
be adjusted to fit results obtained by other methods.

These fundamental discrepancies between the spin excitation spectrum
of the anisotropic $S={1 \over 2}$ two-channel Kondo model and those
of a two channel Fermi liquid indicate that spin-charge
decoupling is an essential feature of two-channel Kondo physics; they
enable us to understand why the spin-$1$ and spin-${1 \over 2}$ models
are so different in the vicinity of the isotropic point.  We have seen
how the triplet and singlet spin excitations take advantage of
spin-charge decoupling to form quasiparticles that are delocalized
between channels in a fashion that does not occur for rigidly defined
electron quasiparticles.  In this respect, the anisotropic two-channel
spin-$1 \over 2$ Kondo model is reminiscent of the Luttinger liquid:
although it displays similar thermodynamics to a Fermi liquid, its
spin-charge decoupled excitations can not be recast in Fermi liquid
form.
This result is of potential practical
importance in systems governed by single-impurity two-channel Kondo
physics, for it ensures the survival of non-Fermi liquid behavior even
when channel asymmetry is destroyed.  Equally important, our results
refute the conventional wisdom, demonstrating that a relevant
perturbation to a non-Fermi liquid state does not inevitably restore
an electron Fermi liquid.

We acknowledge many informative discussions with N. Andrei, A.
Jerez and A. Tsvelik and would like to thank Ph. Nozi\` eres for his
comments on an early version of this work. 
This work was supported by NSF grant DMR-93-12138. A.J.S.
is supported by a Royal Society NATO post-doctoral fellowship.

\end{document}